\documentclass{svjour2}                    
\smartqed  
\usepackage{graphicx}
\usepackage{color}

\usepackage{amssymb}
\usepackage{amsmath}
\usepackage{url}
\newcommand{\ignore}[1]{}

\journalname{Journal of Statistical Physics}
\begin{document}

\title{Searchability of central nodes in networks
\thanks{The author acknowledges funding from Volkswagenstiftung.}
}
\subtitle{}


\author{Konstantin Klemm}


\institute{K. Klemm \at
              Bioinformatics Group, Institute for Computer Science,
	      Leipzig University, H\"{a}rtelstrasse 16-18, 04107 Leipzig,
	      Germany\\
              Tel.: +49-341-97-16704\\
              Fax:  +49-341-97-16679\\
              \email{klemm@bioinf.uni-leipzig.de} 
}

\date{Received: date / Accepted: date}

\maketitle

\begin{abstract}

Social networks are discrete systems with a large amount of
heterogeneity among nodes (individuals). Measures of
centrality aim at a quantification of nodes' importance for
structure and function. Here we ask to which extent the most
central nodes can be found by purely local search. We find
that many networks have close-to-optimal searchability under
eigenvector centrality, outperforming searches for degree and
betweenness. Searchability of the strongest spreaders in
epidemic dynamics tends to be substantially larger for
supercritical than for subcritical spreading.

\end{abstract}

\section{Introduction}


The science of networks \cite{Newman:2010} is an interdisciplinary
enterprise with application areas ranging from systems biology \cite{Alon:2006}
and systems chemistry \cite{Nitschke:2009} to ecology \cite{Bascompte:2007} and
the social sciences \cite{Wasserman:1994}. Methods and models have been
traditionally developed in the fields of graph theory \cite{Diestel:2010} and
algorithms \cite{Thulasiraman:1991,Cormen:2001}.  Dealing with large disordered
and heterogeneous systems, statistical physics now has significant influence
on network science \cite{Albert:2002,Dorogovtsev:2010}. The analysis of
networks further benefits from the growing connections between statistical
physics and computational complexity \cite{Moore:2011}.


On the empirical side, advances in technology for observation and measurement
of large systems and for compilation and storage of the resulting data sets
have triggered a revolution. In the social sciences, interaction networks
typically comprising less than 100 individuals used to be obtained by surveys
\cite{Wasserman:1994} or by direct long-term observation of social groups,
e.g.\ \cite{Zachary:1977}. Nowadays digital communication by mobile phones
\cite{Guimera:2003,Onnela:2007} and e-mail  \cite{Guimera:2003,Onnela:2007} and
through web-based social media such as {\em Facebook}, {\em Twitter} and
{\em Flickr} \cite{Grabowicz:2012} automatically generates data sets covering
millions of individuals.


An important property of social networks is searchability, i.e. the possibility
of finding a target node by iteratively following connections (edges) and  using
information on properties of nodes in the direct neighborhood for deciding which
connection to follow \cite{Adamic:2001,Kleinberg:2000}. This locally available
information, such as the geographical location of an individual and other
personal attributes, may be represented by placing nodes in a Euclidean space
\cite{Boguna:2007a}. A searcher then iteratively chooses the edge that takes it
closest to the target. The success and efficiency of such network navigation
crucially depends on the distribution of long- and short-range connections in
the network \cite{Kleinberg:2000}. Social experiments indicate that these
conditions are fulfilled because personal messages tend to reach the addressee
in a few steps when iteratively forwarded by one person to a chosen acquaintance
\cite{Milgram:1967,Travers:1969,Dodds:2003}. An information theoretic approach
characterizes the reachability between nodes by the minimal description length
of the search path \cite{Rosvall:2005,Sneppen:2005}.


A related task of large practical relevance is the search for the most important
nodes in a given context. These are, for instance, those that would contribute
most to the spreading of an epidemic contagion and thus should be vaccinated
first \cite{Pastor-Satorras:2001,Cohen:2001}. In spreading and other dynamical
processes, the importance of a node strongly depends on the network context
\cite{Kitsak:2010,Klemm:2012}. It may be captured by {\em centrality} measures
such as the number of neighbours, the number of  shortest communication paths
running through, or the number of walks emanating from the node
\cite{Wasserman:1994,Koschutzki:2005}. Given a centrality measure, a {\em
landscape} is obtained by labeling each node with its centrality value
\cite{Wuchty:2003,Axelsen:2006}.

Here we investigate which network structures and centrality measures give rise
to searchable landscapes where the most central nodes are eventually reached by
iteratively jumping to the most central neighbour, i.e.\ by local search. We
quantitatively answer this question by introducing the {\em smoothness} as the
expectation value of the centrality eventually encountered, normalized by the
maximum centrality in the whole network.

In practical scenarios, local search can be applied efficiently only when the
centrality of a node can be obtained locally at that node. Eigenvector
centrality, for instance, is of global nature. The explicit and exact
computation of eigenvector centrality necessarily involves the whole network and
then yields the centrality values of all nodes. Once all these values are known,
the maximum is readily identified. In this case there is no need for local
search on the centrality landscape.

Local search is useful when the network as a whole is not known but centrality
of an encountered node can be obtained indirectly, e.g.\ by local measurement of
system dynamics. The degree of a node is readily obtained by measurement of the
local density from diffusion. The stationary fraction of random walkers at a
node is proportional to the node's number of neighbours. Eigenvector centrality
is measured approximately as the frequency with which a node is involved in
critical spreading or percolation clusters \cite{Klemm:2012}.  We term this
node's property the spreading centrality. The smoothness of the corresponding
landscape quantifies the success of identifying the globally strongest spreaders
by iteratively building a path towards stronger spreaders.

\section{Definitions and Methods}

\subsection{Networks and landscapes}

A network (also called a graph) is a tuple $G=(V,E)$ with $V$ a finite set and
$E$ a set of unordered tuples in $V$. The elements of $V$ are called nodes,
elements of $E$ are called edges. For a node $v \in V$ of a network $(V,E)$,
the neighbourhood $N(v)$ is the set $\{w \in V: \{v,w\} \in E \}$. The
closed neighbourhood of $v$ is $\bar{N}(v)= N(x) \cup \{v\}$.

If $G=(V,E)$ is a network and
$f:V \rightarrow \mathbb{R}$ is an arbitrary mapping, then $L=(V,E,f)$ is called
a landscape (over $G$). The max-neighbourhood of a node $v$ is
the set $N_{\max}(v) = \{w \in N(v) : f(w) = \max_{x \in N(v)} f(x) \}$.
A node $v \in V$ is a [strict] local maximum of the landscape if, for 
all $w\in N(v)$, $f(v) \ge f(w)$ [$f(v) > f(w)$].

\subsection{Search dynamics and smoothness}

An adaptive walk is a local search dynamics. At each time step $t$, a transition
from the current node $v$ to a neighbour is proposed by drawing $w \in N(v)$
uniformly. If $f(w) \ge f(v)$, $w$ is accepted
as the node for the next time step, otherwise the search stays at
$v$. Formally, an adaptive walk on $G$ is a homogeneous (time-independent)
Markov chain with state set $V$ and transition probability
\begin{equation}
\pi(v \rightarrow w) = \left\{ \begin{array}{ll}
\frac{1}{d(v)} & \text{if } w \in N(v) \text{ and } f(w)\ge f(v)\\
0              & \text{otherwise}
\end{array}\right.
\end{equation}
for $w \neq v$, and $\pi(v\rightarrow v)$ given by normalization of
probabilities. An adaptive walk is equivalent to kinetics at zero
temperature where $-f$ plays the role of energy and the network
encodes the allowed transitions between configurations (nodes).
If $v$ is a strict local maximum of $L$, then $\pi(v\rightarrow v) =1$,
so $v$ is an absorbing state for the adaptive walk.

A gradient walk is similar to an adaptive walk, but concentrating on $f$-maximal
neighbours. The dynamics proceeds
to a neighbour with the maximal value for $f$ in the neighbourhood, provided
that this maximal value is strictly larger than the one at the current node.
A gradient walk is thus a Markov chain with transition probability
\begin{equation}
\pi(v \rightarrow w) = \left\{ \begin{array}{ll}
\frac{1}{N_{\max}(v)} & \text{if } w \in N_{\max}(v) \text{ and } f(w)>f(v)\\
0                   & \text{otherwise}
\end{array}\right.
\end{equation}
for $w \neq v$, and $\pi(v\rightarrow v)$ given by normalization. A node $v$
is an absorbing state of the gradient walk ($\pi(v\rightarrow v) =1$), if 
and only if $v$ is a local maximum of the landscape.

Now we characterize the landscape by the success of the search dynamics.
By $\langle f \rangle(t)$ we denote the expectation value of $f$ under a
given search dynamics with the uniform distribution on the node
set $V$ as initial condition. For finite time $t$, we measure the
{\em $t$-smoothness} of $L$,
\begin{equation} \label{eq:smoothfinite}
s_t(L) = \langle f \rangle (t) / \max_{v \in V} f(v)
\end{equation}
to quantify how close the search dynamics approaches nodes with maximum
centrality. The long-term success of the search is quantified by
\begin{equation} \label{eq:smoothdef}
s(L) = \lim_{t\rightarrow\infty} s_t(L)~,
\end{equation}
the {\em smoothness} of $L$. The limit in Eq.~(\ref{eq:smoothdef}) exists
because none of the search steps decreases the $f$-value (so this holds also
for the expectation value) and $f$ is of finite support, thus upper-bounded.
For non-negative functions $f$ --- such as the node centralities considered in
the following --- the
smoothness takes values in the unit interval. In particular, $s(L)=1$ indicates
that the search is certain to reach a node with maximum $f$-value,
regardless of initial condition.

The opposite case, $s(L)<1$, occurs exactly when the Markov chain has an ergodic
set including a node on which $f$ is not maximal. For the gradient walk, this
means that there is a local but not global maximum of $f$ because here the
ergodic sets are exactly the singletons formed by local maxima. In the adaptive
walk, each strict local maximum $v^\ast$ gives rise to an ergodic set
$\{v^\ast\}$; further ergodic sets are formed by interconnected nodes of the
same centrality value, all without neighbours of larger centrality.

\subsection{Centrality measures}

The {\em degree} centrality of a node $v$ is defined as the number of
neighbours
\begin{equation}
d(v) = | N(v)| ~.
\end{equation}
The degree is a purely local measure of node importance, taking into account
only the neighbouring nodes of the node considered. The degree distribution
$P:\mathbb{N}\cup\{0\} \rightarrow \mathbb{R}$ is defined by
\begin{equation}
P(k)= \frac{|\{v \in V : d(v)=k \}|} {|V|}~.
\end{equation}

The $k$-core of a network $G$ is the largest subnetwork of $G$ in which all
nodes have degree at least $k$. It is computed by iteratively removing nodes
with less than $k$ neighbours until no such nodes remain in the network. The
removal of a node deletes all its edges so the loss of node $v$ may
suppress the degree of the neighbours of $v$ below $k$ as well. Therefore the
node removal is done iteratively.

The {\em shell index} $h(v)$ of node $v$ is the largest number $k$ such that
$v$ is in the $k$-core of the network. A large shell index indicates that the
node is part of a densely connected subnetwork.

The {\em eigenvector centrality} is obtained by finding
a non-negative centrality value $e(v)$ for each $v \in V$ to solve the set
of equations
\begin{equation} \label{eq:def_eigenv}
\lambda_\text{max} e(v) = \sum_{w \in N(v)} e(w)
\end{equation}
with $\lambda_\text{max} \in \mathbb{R}$ the maximum value for which such a solution
exists. The underlying idea is that the importance of a node is obtained
self-consistently as the sum of its neighbours' importances discounted by a
node-independent factor $\lambda_\text{max}^{-1}$. Equation~(\ref{eq:def_eigenv}) can be
written  as the eigenvector equation $\lambda_\text{max} e = Ae$ for matrix $A$, hence the
name. The adjacency matrix $A$ has entry $a_{vw}=1$ when $\{v,w\}$ is an edge,
and $a_{vw}=0$ otherwise. The Perron-Frobenius theorem guarantees that the
solution is unique up to a common scaling factor if the network is connected,
i.e. for any two nodes $v$ and $w$, there is a path between $v$ and $w$.

The {\em betweenness centrality} describes node importance as the property of
being contained in shortest paths between other nodes. We denote
by $\sigma_{uw}$ the number of shortest paths between nodes $u$ and $w$,
by $\sigma_{uw}(v)$ the number of such paths passing through node $v$.
Then
\begin{equation}
b(v) = \sum_{u,w \in V \setminus \{v\}} \frac{\sigma_{uw}(v)}{\sigma_{uw}}
\end{equation}
is the betwenness centrality of node $v$. Computation of betweenness centrality
for all nodes in an unweighted network $(V,E)$ requires ${\cal O}(|V| |E|)$ 
computational time steps and ${\cal O}(|V| + |E|)$ memory.

\subsection{Spreading centrality} \label{subsec:spreadc}

Bond percolation \cite{Grimmett:1999} is a theory for the description
of spreading of material and information in disordered media. On arbitrary
networks, the statistics of clusters generated by bond percolation coincides
with those of the basic Susceptible-Infected-Removed (SIR) model of epidemic
spreading \cite{Newman:2002,Keeling:2005}.

A realization of bond percolation on a network $G=(V,E)$ at parameter $\beta$ is
a network $R=(V,E^\prime)$ obtained as follows. For each edge $e \in E$
independently, include $e \in E^\prime$ with probability $\beta$, omit $e$ with
probability $1-\beta$. Such a realization defines a partition of the node set
$V$ into clusters $C_1,C_2,\dots$. Each node is in exactly one cluster. Two
nodes $v$ and $w$ are in the same cluster if and only if $E^\prime$ contains
edges by which one can walk from $v$ to $w$. By $z_\beta(v)$ we denote the
expected size of the cluster containing $v$ in bond percolation at parameter
$\beta$. The spreading centrality $\phi_\beta(v)$ is obtained by
normalization according to 
\begin{equation}
\phi_\beta(v) = \frac{z_\beta(v) - \bar{z_\beta}} 
                    {\max_{w\in V} z_\beta(w) - \bar{z_\beta}} 
\end{equation}
with the mean value $\bar{z_\beta} = \sum_{w \in V} z_\beta(w) / |V|$.
In simulations, $z_\beta(v)$ is obtained from $10^4$ independent realizations
of bond percolation for each network and value of $\beta$.

For each node $v$, $z_\beta(v)$ monotonically increases with $\beta$.
However, the increase may differ from node to node. Therefore the ranking
of nodes with respect to spreading centrality substantially changes with
varying $\beta$ \cite{Klemm:2012}. Then the landscape of spreading 
centrality $L=(V,E,\phi_\beta)$ is qualitatively different for different
values of $\beta$.

\subsection{Random networks} \label{subsec:randgraphs}

Classic random models of networks are by Gilbert \cite{Gilbert:1959} and
Erd\H{o}s and Renyi \cite{Erdos:1959}. Gilbert's random graph model has two
parameters, being the number of nodes $n \in \mathbb{N}$ and the edge
probability $p \in[0,1]$. Among all graphs on $n$ vertices, each graph
with $\eta$ edges is obtained with probability $p^\eta (1-p)^{n(n-1)/2 -\eta}$.
In other words, a realization of the model decides independently for each of
the $n(n-1)/2$ possible edges if the edge is present (with probability $p$)
or absent (with probability $1-p$). Throughout this contribution we refer
to Gilbert's model as {\em random graph}.

The model by Erd\H{o}s and Renyi is defined as the uniform distribution on all
graphs with $n$ nodes and exactly $N$ edges, $n \in \mathbb{N}$ and $N \in
\{0,n(n-1)/2\}$ being the model parameters. Asymptotically, for $n \rightarrow
\infty$, the models by Gilbert and by Erd\H{o}s and Renyi become arbitrarily
similar when taking $\eta = p n(n-1) / 2$.

Another, more versatile form of statistical ensembles are those with a given
degree sequence. The degree of each of the $n$ nodes is a parameter value of the
ensemble. The ensemble consists of all graphs with nodes having the prescribed
degrees, having uniform probability. Here we use Markov chain Monte Carlo
sampling with edge switching \cite{Rao:1996} in order to randomize networks
under conservation of degrees. In a step of edge switching, two node-disjoint
randomly chosen edges $\{v,w\}$ and $\{x,y\}$ are drawn (proposal) and replaced
by edges $\{v,x\}$ and $\{w,y\}$ (acceptance) unless these edges already exist
(rejection). In order to obtain a randomization of a network with $n$ nodes, we
run the Markov chain until switching has been performed $n^2$ times (counting
only accepted steps). The configuration model \cite{Molloy:1995} is an
alternative method for the same purpose.

\subsection{Stochastic network growth} \label{subsec:stochgrowth}

Let us introduce four procedures for building up networks by iterative
addition of nodes and edges. The following scenario is common to all
these procedures. The network is initialized with $m$ nodes fully
interconnected, i.e. all $m(m-1)/2$ edges are present, where $m$ is a parameter.
Then in each step $t$ of growth, a subset $S$ of size $m$ is
chosen stochastically from the current node set; a new node $v(t)$ and
$m$ new edges are added to the network, where each edge connects one
node in $S$ with $v(t)$. The growth step is iterated until the network
reaches the desired size. Each specific procedure is defined by the stochastic
choice of the set of nodes that the new edges are attached to.

The {\em preferential attachment} rule
\cite{Barabasi:1999,Krapivsky:2000,Dorogovtsev:2000} 
(also termed {\em cumulative advantage} \cite{Price:1976})
adds edges preferably
to nodes having large degree already. Starting with an initially empty set $S$,
a node $v$ is drawn from the distribution
\begin{equation}
\pi(v) = \frac{d(v)+a}{\sum_{w \in V \setminus S} (d(w)+a)}
\end{equation}
on $V\setminus S$. Then this node is included in $S$. This drawing of nodes
without replacement is repeated until $|S|=m$. The parameter $a$ is a
constant bias with $m < a < \infty$. {\em Uniform attachment}, called model A in refs.\
\cite{Barabasi:1999,Barabasi:1999b}, is obtained in the limit
$a \rightarrow \infty$. Here $S(t)$ is drawn uniformly from the set of all
$m$-node subsets.

The {\em edge attachment} rule by Dorogovtsev et al.\ \cite{Dorogovtsev:2001}
is defined specifically for $m=2$ edges to be added per node. Rather than
selecting nodes separately, the set $S$ is determined by drawing an
{\em edge} $e \in E$ uniformly and setting $S=e$. Thus each new
node is attached to the end nodes of an existing edge, thereby forming a
triangle. 

In the {\em deactivation model} by Klemm and Egu\'{\i}luz
\cite{Klemm:2002a,Klemm:2002b,Eguiluz:2002}, the set of $S$ of nodes receiving
edges---thus called active---evolves over time. Upon initialization,
$S$ comprises all $m$ nodes present initially. In each growth each step, after adding
a node $v$ and its $m$ edges, $v$ is included in $S$. From the $|S|=m+1$ then
contained in $S$, one node $u$ is chosen for removal (deactivation) by the
distribution
\begin{equation}
\pi (u) = \frac{d(u)^{-1}}{\sum_{w \in S} d(w)^-1}~.
\end{equation}

The expected degree distribution from the deactivation model and from edge
attachment decay as a power law $P(k) \sim k^{-\gamma}$ with $\gamma \approx 3$.
Preferential attachment yields $P(k) \sim k^{-\gamma}$ with $\gamma = 3 + a/m$.
The asymptotic degree distribution from uniform attachment decays geometrically.
The deactivation model generates networks with average distance between nodes
increasing linearly with size; the other three rules generate networks with
distances increasing logarithmically or sublogarithmically.

Note that all these procedures generate networks with shell index $h(v) = m$
for all nodes $v$, as shown by induction over the number of nodes. The initial
condition is a fully interconnected network of $m$ nodes. After adding a node
and $m$ links, each node $v$ now has exactly $d(v)=m$ neighbours and therefore
$h(v)=m$. After an arbitrary number of node additions, each node has at least
$m$ neighbours so that the $m$-core is the whole network. In the computation
of the $m+1$-core, nodes with degree $m$ or smaller are iteratively removed,
which amounts to disintegrating the whole network in node order similar to its
generation. Thus the $m+1$-core is empty. Each node has the same shell index
$m$, which is the global maximum. It follows trivially that all these networks
have smoothness 1 with respect to shell index.

\section{Numerical results}

\begin{figure}
\centerline{\includegraphics[width=\textwidth]{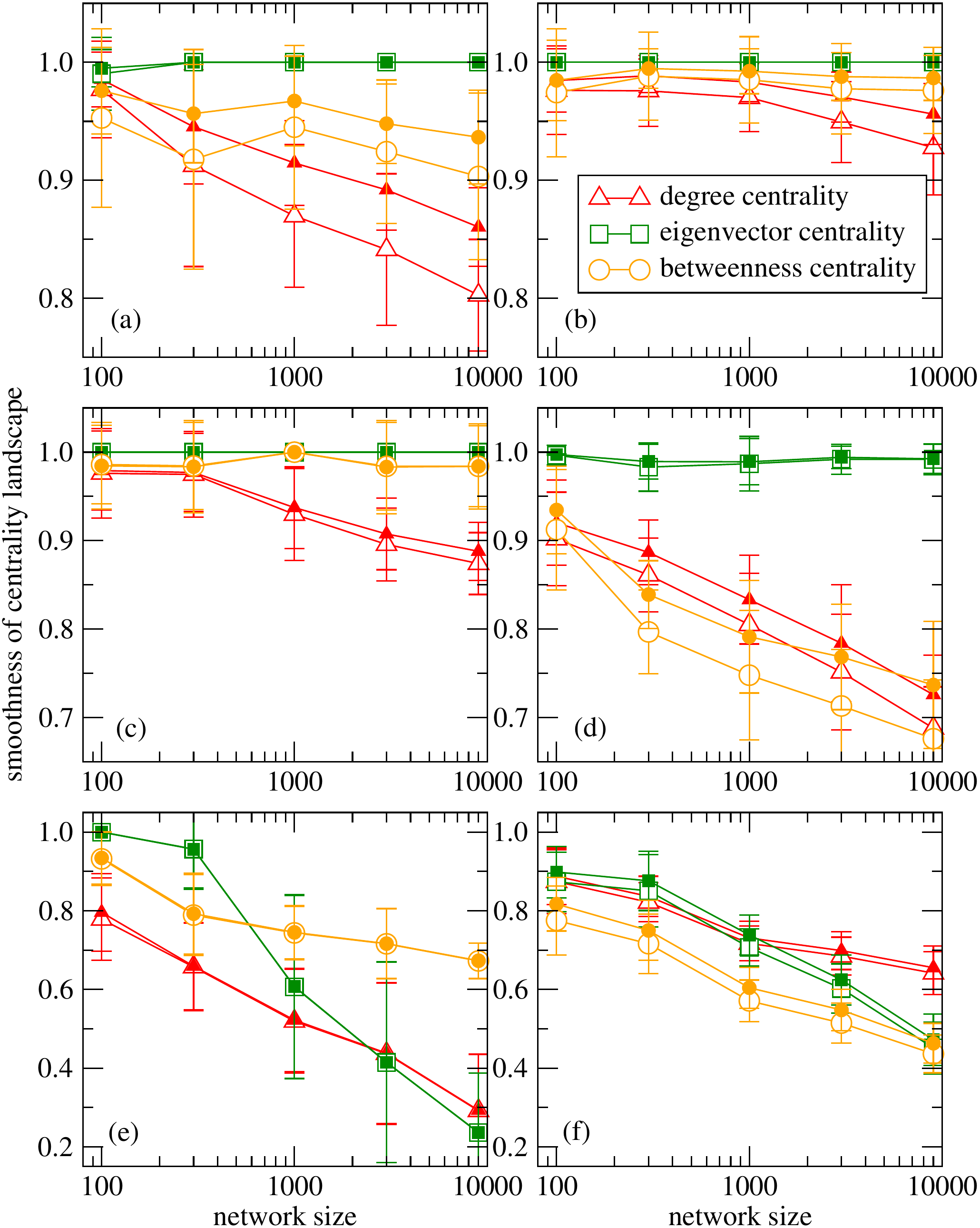}}
\caption{\label{fig:land_0}
Smoothness of centrality landscapes on stochastically grown networks.
Symbols distinguish centrality measures (see legend). Large open symbols
are for search by adaptive walk, small filled symbols for gradient walk.
Panels (a-f) distinguish procedures for network generation.
(a) preferential attachment with bias $a=0$, degree exponent $\gamma =3$
(b) preferential attachment with bias $a=-m/2$, $\gamma =2.5$.
(c) edge attachment, (d) uniform attachment, (e) deactivation model.
(f) random graph with edge probability $p=4/(n-1)$.
Error bars indicate standard deviation over 10 independent realizations
of network generation. In the growing networks (a)-(e), $m=2$ edges per node are attached
in the growth process; in (f) edge probability is $p=4/(n-1)$,
resulting in asymptotic average degree of $4$ in all cases.
}
\end{figure}

We first study smoothness of centrality landscapes for stochastically
generated network instances, as described in sections
~\ref{subsec:randgraphs}~and~\ref{subsec:stochgrowth}.
The dependence of smoothness on the number of nodes is displayed in
Figure~\ref{fig:land_0}. Smoothness depends both on the type of network and the
centrality measure under consideration. Except for the random graphs and the
networks grown by the deactivation model, eigenvector landscapes have a maximum
smoothness $s=1$ with deviations only for small networks. Betweenness centrality
landscapes have smoothness close to 1 in networks grown by edge attachment. For
all other combinations of network generation and centrality measure, a decrease
of smoothness with the size of the network is observed. In those cases,
centrality landscapes become less searchable with increasing size.

The aforementioned results are qualitatively the same for both types of local
search. Except for the networks from uniform attachment, smoothness values from
adaptive and from gradient walks are also quantitatively the same up to
statistically insignificant deviations.

\begin{table}

\caption{\label{tab:netsmooth}
Smoothness of centrality landscapes for six empirical networks.
In each column, the first value gives the smoothness for search by
adaptive walk. The second value is for search by gradient walk.
For each network, the first (upper) row of smoothness values is
for the original network. The second row is for randomized surrogate
networks with the same degree sequence (see section~\ref{subsec:randgraphs});
each value is the mean of the smoothness values of 10 independent
randomizations. Networks are
(a)~the e-mail contacts from University Rovira i Virgili, restricted to the 
largest connected component \cite{Guimera:2003};
(b)~jazz bands connected by an edge if they share a musician
\cite{gleiser:2003};
(c)~authors in the cond-mat e-print archive (arXiv) where $\{v,w\}$ is
an edge if $v$ and $w$ have co-authored a paper \cite{Newman:2001},
updated version including data until March 2005;
(d)~users of the Pretty-Good-Privacy algorithm for secure information
interchange \cite{Boguna:2004};
(e)~Internet at Autnomous Systems level, snapshot taken on July 06, 2006.
(f)~electric power grid with generators, transformers and substations as nodes,
edges being high-voltage transmission lines \cite{Watts:1998}.
Network data (a), (b) and (d) were downloaded from
\protect\url{http://deim.urv.cat/~aarenas/data/welcome.htm} ;
(c), (e) and (f) from
\protect\url{http://www-personal.umich.edu/~mejn/netdata/}
}

\begin{tabular}{|l|l|l|l|l|l|}  \hline
network  & $|V|, |E|$   &      degree & eigenvector & shell index & betw.ness\\ \hline
(a) e-mail   & 1133, 5451   & 0.882 0.949 & 1.000 1.000 & 0.999 1.000 & 0.867 0.908\\
             &              & 0.859 0.892 & 0.905 0.939 & 0.992 0.992 & 0.791 0.851\\
& & & & &\\
(b) jazz     & 198, 2742    & 1.000 1.000 & 1.000 1.000 & 1.000 1.000 & 0.830 0.934\\
             &              & 1.000 1.000 & 1.000 1.000 & 1.000 1.000 & 1.000 1.000\\
& & & & &\\
(c) cond-mat & 40421, 175692& 0.766 0.773 & 0.921 0.921 & 0.791 0.794 & 0.627 0.671\\
             &              & 0.951 0.967 & 0.957 0.971 & 0.995 0.995 & 0.926 0.955\\
& & & & &\\
(d) pgp      & 10680, 24340 & 0.569 0.607 & 1.000 1.000 & 0.848 0.852 & 0.736 0.783\\  
             &              & 0.832 0.853 & 0.861 0.870 & 0.898 0.898 & 0.813 0.846\\
& & & & &\\
(e) Internet & 22963, 48436 & 0.973 0.981 & 1.000 1.000 & 1.000 1.000 & 0.996 0.997\\
             &              & 0.869 0.872 & 0.892 0.892 & 0.895 0.895 & 0.888 0.889\\
& & & & &\\
(f) power grid & 4941, 6594 & 0.401 0.410 & 0.230 0.233 & 0.651 0.651 & 0.411 0.437\\ 
             &              & 0.474 0.495 & 0.623 0.630 & 0.970 0.970 & 0.395 0.426\\ \hline
\end{tabular}
\end{table}

Now we turn to centrality landscapes based on empirical networks. 
Table~\ref{tab:netsmooth} provides an overview of smoothness values for
four social networks (a-d) and two technological networks (e-f) further
described in the caption. We find that eigenvector centrality
induces the smoothest landscapes on the social networks and on the
Internet snapshot (e). On the power grid (f), however,
eigenvector centrality reaches lower smoothness than the other three
centralities. This network is also the minimum of each of the eight smoothness
columns, i.e. the power grid yields the lowest smoothness among the networks
under a given centrality measure and walk type.

Randomization of these networks under conservation of the degree sequence
drastically changes the smoothness values. The searchability of centrality
landscapes depends on properties of the empirical networks beyond the degree
distribution.

\begin{figure}
\centerline{\includegraphics[width=\textwidth]{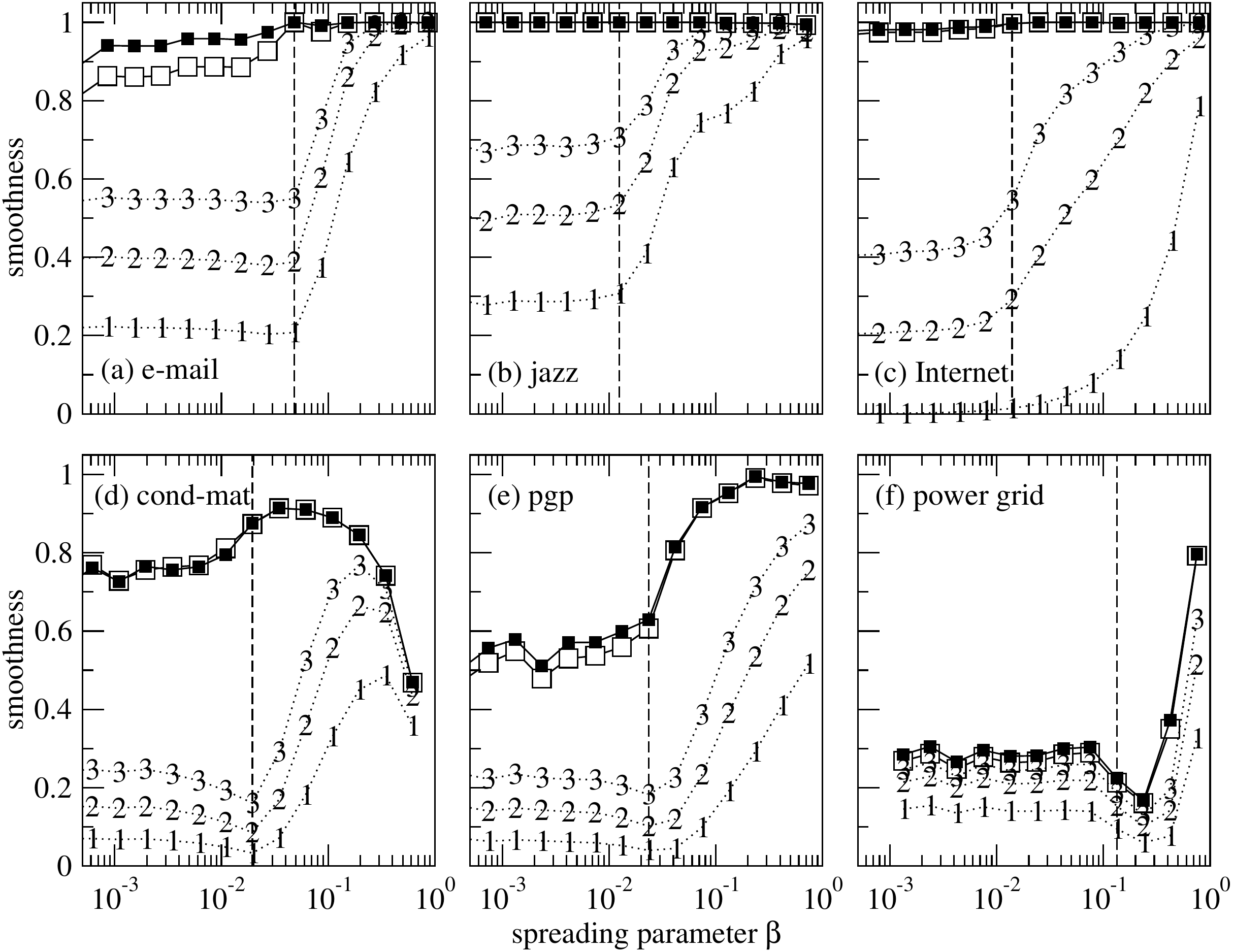}}
\caption{\label{fig:epi_0}
Smoothness of spreading centrality landscapes under adaptive walks (open squares)
and gradient walks (filled squares). 
Vertical dashed lines indicate the parameter value $\beta_c= 1 / \lambda_\text{max}$
as an estimate of the transition point from local to global spreading
\cite{Chakrabarti:2008}. Curves with number symbols $t \in \{1,2,3\}$ show $s_(L)$, the 
{\em $t$-smoothness} of the adaptive walk at time $t$,
cf.\ Equation~(\ref{eq:smoothfinite}).
See the caption of Table~\ref{tab:netsmooth} for details on the
networks.
}
\end{figure}

For landscapes of spreading centrality (see section~\ref{subsec:spreadc}), the
smoothness as a function of the spreading parameter $\beta$ is plotted in
Figure~\ref{fig:epi_0}. For small $\beta$, smoothness values are close to those
of the corresponding degree landscapes, see Table~\ref{tab:netsmooth}.  This
finding is in accordance with the large rank order correlation between spreading
centrality and degree below the transition \cite{Klemm:2012}. In the transition
regime ($\beta\approx \beta_c$), smoothness increases with $\beta$. The rank
order correlation between degree and spreading centrality is also large in the
supercritical regime. Here, however, smoothness for spreading centrality does
not coincide with smoothness for degree. The $\beta$-dependence of smoothness
varies across networks in the supercritical regime.

Above the transition, the $t$-smoothness values $s_t(L)$ for small number of
time steps $t$ strongly increases, indicating an accelerated success of the
search for more central nodes. Typically in the supercritical regime, $t=2$ or
$t=3$ steps are sufficient for the adaptive walk to saturate, i.e. to
reach $s_t(L) \approx s(L)$.

\section{Estimates for a given degree distribution}

Let us estimate the smoothness of degree centrality landscapes on the basis of
the degree distribution alone. To this end, we perform an annealed
approximation. At each time step, the network is drawn uniformly at random from
the set of networks with the given degree distribution. 

From the given degree distribution $P(k)$ we obtain the probability of
encountering a node of degree $k$ by following a uniformly chosen edge as 
\begin{equation}
P^\prime(k) = k P(k) / \bar{k}
\end{equation}
with the average degree $\bar{k} = \sum_{k=0}^\infty k P(k)$.
We use the cumulative of $P^\prime$ being
\begin{equation}
Q^\prime(k) = \sum_{i=k}^\infty P^\prime(k)~.
\end{equation}

We define $x(k,t)$ as the fraction of searches being active at a node of degree
$k$ at time $t$. A search is active at initialization and turns inactive when
reaching a node that is a strict local maximum w.r.t.\ degree. By $y(x,t)$ we
denote the fraction of searches being inactive at a node of degree $k$ at time
$t$. We have $\sum_{k=0}^\infty x(k,t)+y(k,t)=1$ as a normalization at all times
$t$. Initially, all searches are active and distributed uniformly across
nodes, so $x(k,0) = P(k)$ and $y(k,0) =0$ for all $k$. 

If all neighbours of a node with degree $k$ have degree strictly smaller than
$k$, the adaptive walk stays there, so the search becomes inactive.
This happens with probability
\begin{equation}
\alpha(k)=[1-Q^\prime(k)]^k~.
\end{equation}
Thus we have
\begin{equation} \label{eq:ratey}
y(k,t+1) = y(k,t) + \alpha(k) x(k,t)
\end{equation}
Otherwise, the searcher eventually finds a neighbouring node with degree at
least $k$ and jumps there. In order to simplify the equations, we assume that
this jump happens immediately in one time step. Compared to the real dynamics,
where one or several rejections may occur before the jump, this simplification
modifies the transients but not the asymptotic distribution of searches across
degrees. When following a uniformly chosen edge, we find a node of degree $k$
with probability $P^\prime(k) / Q^\prime(l)$ under the constraint $k \ge l$.
Therefore the active searches propagate as
\begin{equation} \label{eq:ratex}
x(k,t+1) = \sum_{l=1}^{k} (1-\alpha(l)) [P^\prime(k) / Q^\prime(l)] x(l,t)~.
\end{equation}
When iterating Equations~(\ref{eq:ratey},~\ref{eq:ratey}) over time steps,
the fraction of active searches converges to zero. The estimate
$s^\ast(L)$ of the smoothness is obtained from the asymptotic distribution
of inactive searches
\begin{equation}
s^\ast(L) = \lim_{t \rightarrow \infty}
\sum_{k=0}^\infty y(k,t) k / k_\text{max}
\end{equation}
with the maximum degree $k_\text{max}$.

\begin{figure}
\centerline{\includegraphics[width=\textwidth]{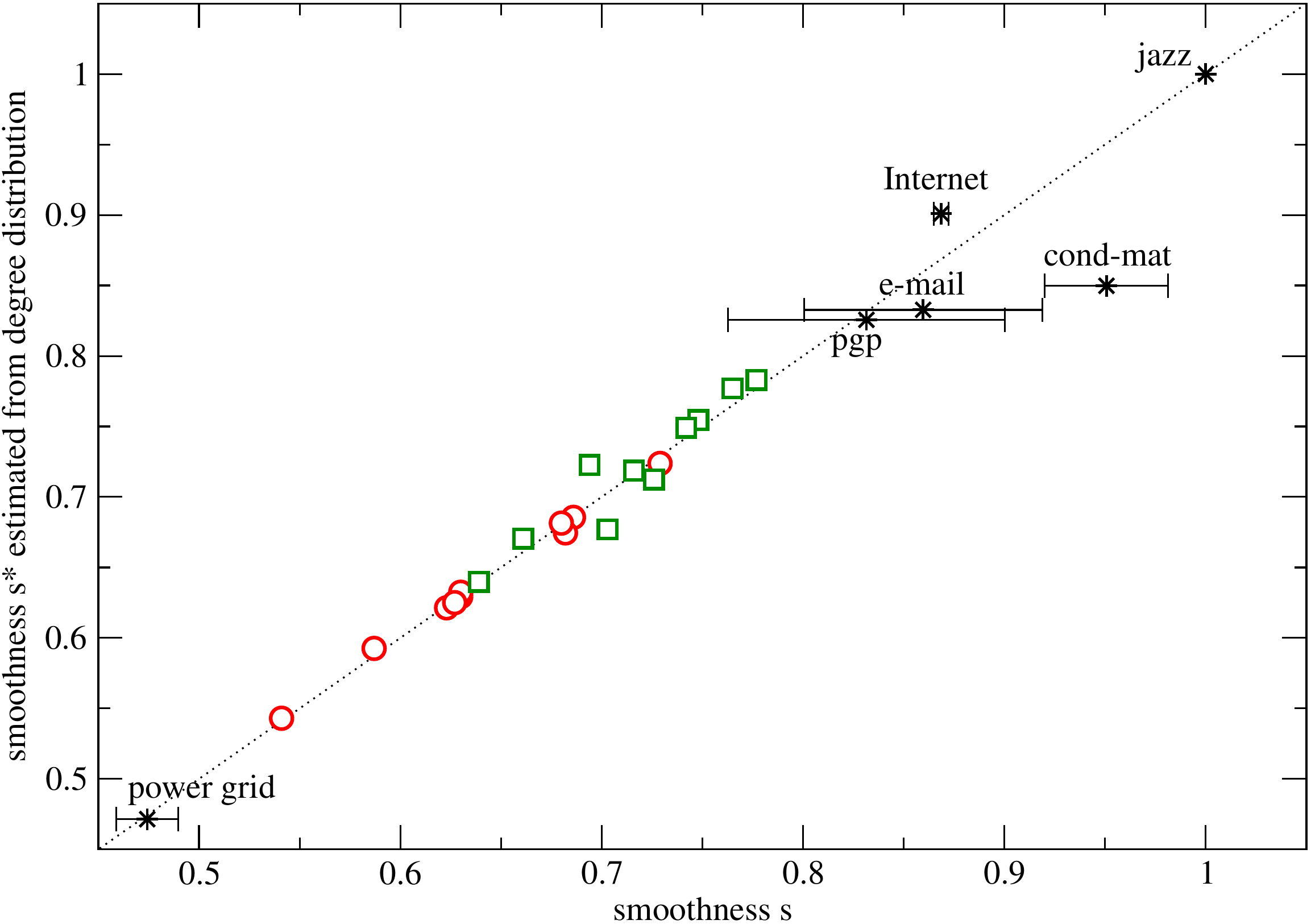}}
\caption{\label{fig:simconf_0}
Smoothness estimates $s^\ast$ compared to the real smoothness for
adaptive walks on degree centrality landscapes.
Symbols distinguish networks as follows.
Squares: 10 independent realizations of random graphs with $n=1000$ nodes
and parameter $p=4/(n-1)$. Cycles: Same as for squares but $n=9000$.
Stars: empirical networks randomized by switching under conserved degrees,
cf.\ subsection~\ref{subsec:randgraphs}. The abscissa of each star is the
smoothness averaged over 10 independent realizations of randomization
(see also Table~\ref{tab:netsmooth}). Error bars indicate 
the standard deviation over the 10 realizations. 
}
\end{figure}

Figure~\ref{fig:simconf_0} provides a comparison between actual smoothness value
$s$ and the estimates $s^\ast$ for several networks. For realizations of random
graphs, the degree distribution provides almost complete information on
smoothness. The typical difference in smoothness $s$ across realizations is
significantly larger than the deviation of the estimate $s^\ast$ from the true
value. Applied to randomizations of empirical networks, there is reasonable
agreement between $s$ and $s^\ast$ in most of the cases.

\section{Discussion and outlook}

We have defined a framework for assessing the searchability of
centrality landscapes arising from a network. The {\em smoothness} of a
network combined with a centrality quantifies the extent to which nodes
of large centrality are found by searching the network locally. Local
search means that the sequence of nodes visited is a
walk following edges on the network.

This local perspective is motivated by the limited information on the network.
Rather than knowing the whole network at the outset, a searching agent explores
the system by iteratively following connections. In the present framework, we
have assumed that the centrality value of each node becomes available ad hoc
when encountering the node. This is the case when the centrality values
themselves involve local information only, e.g.\ the degree. Alternatively,
the centrality of interest results from local measurement on dynamics taking
place on the network. We have investigated the spreading centrality as an
example.

Analytic insight into smoothness $s$ of centrality landscapes is
desirable, e.g.\ lower bounds on $s$ depending on network properties. As a
step in this direction, we have defined rate equations for estimating
smoothness of degree landscapes in an annealed approximation.

Numerically we find that eigenvector centrality typically generates maximally
($s=1$) or almost maximally smooth landscapes. As a heuristic explanation,
smoothness arises due to the summation in the eigenvector equations that amounts
to an averaging over neighbouring nodes. On the stochastically generated
networks lacking small-world property \cite{Klemm:2002b} and the extremely
sparse electric power grid, none of the centrality measures under consideration
generates smooth landscapes. Small neighbourhoods  and long distances in search
space tend to create obstacles to local search.

\bibliographystyle{spmpsci}      
\bibliography{epiwalks}   

\end{document}